\documentclass[aps,pra,preprint,superscriptaddress,longbibliography]{revtex4-2}

\usepackage{amsmath,amsthm,amsfonts,amssymb}
\usepackage{bm}
\usepackage{xcolor,graphicx}
\usepackage[pdftex,colorlinks=true,linkcolor=blue,citecolor=blue,urlcolor=blue]{hyperref}

\begin{document}

\title{Classification of Two-Mode Laguerre--Gaussian Polarization Textures}

\author{{L.~E.} {Lara Rosales}}
\email[e-mail: ]{leslie.lara@cio.mx}
\affiliation{Centro de Investigaciones en Óptica, A.C., Loma del Bosque 115, Colonia Lomas del Campestre, 37150 León, Gto., Mexico.}

\author{{M. F.} {Ferrer-Garcia}}
\email[e-mail: ]{mferrerg@uottawa.ca}
\affiliation{Nexus for Quantum Technologies, Department of Physics, University of Ottawa, Ottawa, K1N 6N5, ON, Canada}

\author{{B.~M.} {Rodr\'iguez-Lara}} 
\email[e-mail: ]{blas.rodriguez@gmail.com}
\affiliation{Universidad Polit\'ecnica Metropolitana de Hidalgo, Tolcayuca, Hidalgo 43860, Mexico.}

\date{\today}

\begin{abstract}
Polarization textures shape the local polarization across a beam and support optical communication, imaging, metrology, and light--matter interaction.
Their charge classifies them only partially, as textures with equal charge can display different ring patterns.
We classify textures of two equally weighted Laguerre--Gaussian modes with opposite circular polarizations, arbitrary indices, and a common waist by relating the mode indices to the polarization at the beam center and far from it.
If the polarization near the center or far away depends on direction, the charge can be noninteger.
We find where the polarization becomes purely circular or the light vanishes, and calculate how the regions between those rings contribute to the charge.
As the beam propagates, the transverse polarization pattern expands and the polarization can rotate, while the charge of each region stays the same.
In one family, we find two more rings with canceling contributions whenever each mode has one more radial node, while the total charge stays unchanged.
Mode pairs with the same polarization at the center and far away, and the same charge, can still rotate by different amounts.
Our results may guide the design of beams with prescribed polarization rings and charge.
\end{abstract}

\maketitle
\clearpage

\section{Introduction}
\label{sec:Sec1}

Structured light encompasses optical fields with spatially varying amplitude, phase, or polarization.
The spatial distribution of local polarization defines a polarization texture and maps the transverse plane to the Poincar\'e sphere~\cite{Shen2025p295}.
Advances in beam shaping have enabled the generation of fields with tailored polarization textures, including vector vortex beams~\cite{Milione2011p053601,Ndagano2018p292} and full Poincar\'e beams~\cite{Beckley2010p10777,Beckley2012p9357}.
These fields can exhibit polarization singularities~\cite{Liu2021p1469} and spin--orbit coupling~\cite{Bliokh2015p796}, while their spatial structure supports vectorial field control~\cite{Wu2022p187}.
Control of the local polarization supports optical communication~\cite{Milione2015p4887,Zhao2015p4843,Sit2017p1006}, imaging and microscopy~\cite{Kozawa2018p86,Cheng2022p2780}, precision metrology~\cite{Fang2021p4186,QuintoSu2023p3693}, optical trapping~\cite{Bhebhe2018p6255892,Yang2021p034001}, and light--matter interaction~\cite{Wang2020p031702,Verde2023p21283,RosalesGuzman2024p200503}.

The normalized Stokes maps of these polarization textures cover regions of the Poincar\'e sphere, and the local orientation of each map determines the sign of each contribution.
The integrated skyrmion charge $Q$ is the sum of the signed areas of those regions, counting repeated coverage, divided by the sphere's total area~\cite{McWilliam2023p23030155}.
Contributions from oppositely oriented regions can cancel, while repeated coverage with the same orientation adds.
When the polarization map extends continuously across the transverse plane and approaches one polarization state in every direction at large radius, the integrated skyrmion charge equals the integer degree of the extended map~\cite{Nakahara2018,Manton2010}.
For an axially symmetric texture, the polarization limits at the core and at large radius define its boundary class; continuity requires a unique core limit independent of the direction of approach.
Noninteger integrated charges can arise when the core limit depends on the direction of approach~\cite{Ye2024p240411530}, or in superpositions with fractional orbital angular momentum~\cite{Cao2026p063537}.

Superpositions of orthogonally polarized Laguerre--Gaussian modes generate higher-order Poincar\'e beams, some of which support optical skyrmions~\cite{Milione2011p053601,Gao2020p053513,Gao2021p049901}.
Rational maps provide a way to construct optical skyrmions~\cite{Cisowski2023p2200350}.
More generally, a non-Abelian Chern--Simons gauge theory describes the nonseparability of spatial modes and polarization in vector beams~\cite{Ornigotti2026p033503}.
Nodes of the constituent modes can divide the texture into concentric regions~\cite{Galvez2012p2925}, whose charges may cancel even as the number of rings changes, as illustrated by skyrmionium and target textures~\cite{Shen2024p024025}.
For modes with unequal beam widths, the slower-decaying mode determines the polarization at large radius, and the charge can change if the widths cross during propagation~\cite{Gao2020p053513,Ye2024p240411530}.
With a common waist, the Gaussian factors cancel, yet textures with the same total charge can have different polarization rings.

We classify polarization textures formed by two equally weighted Laguerre--Gaussian modes with opposite circular polarizations and a common waist, connecting their mode indices to boundary polarization, ring patterns, charge, and free propagation.
We construct their fields in Sec.~\ref{sec:Sec2} and show that distinguishing textures with the same total charge requires both the leading radial powers, which determine nine boundary classes in Sec.~\ref{sec:Sec3}, and the finite-radius nodes, which determine pole and dark rings in Sec.~\ref{sec:Sec4}.
At nonzero Stokes winding, textures with an intermediate core or asymptotic latitude are open and can have noninteger integrated charge; two single-valued boundary limits define an integer degree.
We calculate the annular charges in Sec.~\ref{sec:Sec4}; for left-circular modes of azimuthal index zero and right-circular modes of index minus one, each increment in the shared radial index adds two pole rings with canceling charges.
We show that free propagation preserves the annular and full-plane charges while the relative Gouy phase rotates the Stokes vectors in Sec.~\ref{sec:Sec5} and compare five boundary classes with distinct ring sequences in Sec.~\ref{sec:Sec6}.
In Sec.~\ref{sec:Sec7}, we summarize the results and discuss their implications.

\section{Two-mode Laguerre--Gaussian polarization family}
\label{sec:Sec2}

We use the real electric field $\operatorname{Re}\left[ \bm{\mathcal{E}} \left( \bm{r}, z \right) e^{i \left( k z - \omega t \right)} \right]$ for forward propagation, with transverse position $\bm{r} = (x, y)$, longitudinal coordinate $z$, time $t$, wavenumber $k = 2 \pi / \lambda$ for wavelength $\lambda$, and angular frequency $\omega$.
We construct its envelope from two normalized Laguerre--Gaussian modes with equal weights and opposite circular polarizations,
\begin{align}
	\begin{aligned}
		\bm{\mathcal{E}} \left( \bm{r}, z \right) =&~ \frac{1}{\sqrt{2}} \left[ \Psi_{p_{\mathrm{L}}, \ell_{\mathrm{L}}} \left( \bm{r}, z \right) \bm{e}_{\mathrm{L}} \right. \\
		& \left. + e^{-i \gamma} \Psi_{p_{\mathrm{R}}, \ell_{\mathrm{R}}} \left( \bm{r}, z \right) \bm{e}_{\mathrm{R}} \right],
	\end{aligned}
\end{align}
with radial indices $p_{\mathrm{L}}, p_{\mathrm{R}} = 0, 1, 2, \ldots$, azimuthal indices $\ell_{\mathrm{L}}, \ell_{\mathrm{R}} = 0, \pm 1, \pm 2, \ldots$, and constant relative phase $-\gamma$.
We use the left-circular unit vector $\bm{e}_{\mathrm{L}} = (\bm{e}_{x} + i \bm{e}_{y}) / \sqrt{2}$ and the right-circular unit vector $\bm{e}_{\mathrm{R}} = (\bm{e}_{x} - i \bm{e}_{y}) / \sqrt{2}$, with handedness defined for an observer facing the source.
In the transverse plane, $\bm{e}_{x}$ is the unit vector along the horizontal $x$ axis and $\bm{e}_{y}$ along the vertical $y$ axis.
The opposite circular polarizations make the two vector modes orthogonal for any choice of scalar mode indices.
Our orthonormal Laguerre--Gaussian modes $\mathrm{LG}_{p,\ell}$~\cite{MoralesRodriguez2024p033523,AguirreOlivas2025p1121},
\begin{align}
	\Psi_{p, \ell} \left( \bm{r}, z \right) = \frac{\sqrt{2}}{w \left( z \right)} e^{\frac{i k r^{2}}{2 R \left( z \right)}} e^{-i N_{p, \ell} \varPhi \left( z \right)} \psi_{p, \ell} \left( \rho, \phi \right),
\end{align}
share the waist size $w_{0}$, waist plane $z = 0$, and wavelength $\lambda$, with Rayleigh range $z_{R} = \pi w_{0}^{2} / \lambda$.
We use the beam width $w(z) = w_{0} \sqrt{1 + (z / z_{R})^{2}}$, inverse curvature radius $1 / R(z) = z / (z^{2} + z_{R}^{2})$, and Gouy phase $\varPhi(z) = \arctan(z / z_{R})$.
We use the azimuth $\phi$ and the scaled radius $\rho = \sqrt{2} r / w(z)$, with radial coordinate $r = \lvert \bm{r} \rvert$.
The total mode number
\begin{align}
	N_{p, \ell} = 2 p + \lvert \ell \rvert + 1
\end{align}
multiplies the Gouy phase $\varPhi(z)$ in the propagation factor.
The transverse profile
\begin{align}
	\begin{aligned}
		\psi_{p, \ell} \left( \rho, \phi \right) =&~ \left( -1 \right)^{p} \sqrt{\frac{p!}{\pi \left( p + \lvert \ell \rvert \right)!}} \rho^{\lvert \ell \rvert} e^{-\frac{1}{2} \rho^{2}} \\
		& \times L_{p}^{\lvert \ell \rvert} \left( \rho^{2} \right) e^{i \ell \phi}
	\end{aligned}
\end{align}
uses the generalized Laguerre polynomial $L_{p}^{\lvert\ell\rvert}$ of degree $p$.
The mode has $p$ radial nodes, phase winding $\ell$, and axial orbital angular momentum $\ell \hbar$ per photon.
The beams are orthonormal under the transverse inner product,
\begin{align}
	\int \mathrm{d}^{2} \bm{r} \, \Psi_{p^{\prime}, \ell^{\prime}}^{\ast} \left( \bm{r}, z \right) \Psi_{p, \ell} \left( \bm{r}, z \right) = \delta_{p^{\prime}, p} \delta_{\ell^{\prime}, \ell},
\end{align}
at every propagation plane $z$.

\begin{figure}[t]
    \centering
    \includegraphics[width=0.75 \textwidth]{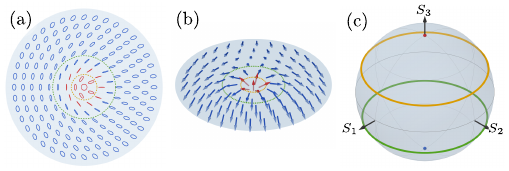}
    \caption{Polarization texture of left-circular $\mathrm{LG}_{0,0}$ and right-circular $\mathrm{LG}_{0,1}$ modes at $z=0$ and $\gamma=0$.
    (a) Polarization ellipses and (b) normalized Stokes vectors across the transverse plane, colored red for left-handed polarization, blue for right-handed polarization, and black for linearly polarized states, and (c) the trajectories of the gold inner and green outer circles on the Poincar\'e sphere together with its two poles.}
    \label{fig:Fig1}
\end{figure}

We use the spatially resolved Stokes parameters~\cite{McWilliam2023p23030155},
\begin{align}
	\begin{aligned}
		S_{0} =&~ \frac{1}{2} \left( \lvert \Psi_{p_{\mathrm{L}}, \ell_{\mathrm{L}}} \rvert^{2} + \lvert \Psi_{p_{\mathrm{R}}, \ell_{\mathrm{R}}} \rvert^{2} \right), \\
		S_{1} =&~ \operatorname{Re} \left[ e^{-i \gamma} \Psi_{p_{\mathrm{R}}, \ell_{\mathrm{R}}} \Psi_{p_{\mathrm{L}}, \ell_{\mathrm{L}}}^{\ast} \right], \\
		S_{2} =&~ \operatorname{Im} \left[ e^{-i \gamma} \Psi_{p_{\mathrm{R}}, \ell_{\mathrm{R}}} \Psi_{p_{\mathrm{L}}, \ell_{\mathrm{L}}}^{\ast} \right], \\
		S_{3} =&~ \frac{1}{2} \left( \lvert \Psi_{p_{\mathrm{L}}, \ell_{\mathrm{L}}} \rvert^{2} - \lvert \Psi_{p_{\mathrm{R}}, \ell_{\mathrm{R}}} \rvert^{2} \right),
	\end{aligned}
\end{align}
to describe the local polarization.
At points where the field is nonzero, with intensity $S_{0} > 0$, the normalized Stokes vector $\bm{s} \left( \bm{r}, z \right) = \left( S_{1}, S_{2}, S_{3} \right) / S_{0}$ has unit length and represents a point on the Poincar\'e sphere.
Left-circular polarization corresponds to the north pole $\bm{s} = (0, 0, +1)$ of the Poincar\'e sphere, and right-circular polarization corresponds to the south pole $\bm{s} = (0, 0, -1)$.
At each propagation plane $z$, the normalized Stokes vector field is the polarization texture,
\begin{align}
	\bm{s} \left( \bm{r}, z \right) = \frac{1}{1 + \lvert \mu \rvert^{2}} \left( 2 \operatorname{Re} \left[ \mu \right], 2 \operatorname{Im} \left[ \mu \right], 1 - \lvert \mu \rvert^{2} \right),
\end{align}
with the modal ratio as its stereographic coordinate,
\begin{align}
	\mu \left( \bm{r}, z \right) = e^{-i \gamma} \frac{\Psi_{p_{\mathrm{R}}, \ell_{\mathrm{R}}} \left( \bm{r}, z \right)}{\Psi_{p_{\mathrm{L}}, \ell_{\mathrm{L}}} \left( \bm{r}, z \right)},
\end{align}
with zero modulus at the north pole and divergent modulus at the south pole.
The local polarization latitude depends on the modulus of the modal ratio, and the in-plane orientation on its phase.
We show the polarization ellipses in Fig.~\ref{fig:Fig1}(a), the corresponding normalized Stokes vectors in Fig.~\ref{fig:Fig1}(b), and the trajectories of the gold inner and green outer transverse circles on the Poincar\'e sphere in Fig.~\ref{fig:Fig1}(c).

We cancel the common Gaussian envelope and curvature phase to obtain the modal ratio
\begin{align}
	\mu \left( \bm{r}, z \right) = A \left( \rho \right) e^{i \left[ m \phi + \Gamma \left( z \right) \right]},
\end{align}
with the signed radial factor
\begin{align}
	\begin{aligned}
		A \left( \rho \right) =&~ \sqrt{\frac{p_{\mathrm{R}}! \left( p_{\mathrm{L}} + \lvert \ell_{\mathrm{L}} \rvert \right)!}{p_{\mathrm{L}}! \left( p_{\mathrm{R}} + \lvert \ell_{\mathrm{R}} \rvert \right)!}} \\
		& \times \rho^{\lvert \ell_{\mathrm{R}} \rvert - \lvert \ell_{\mathrm{L}} \rvert} \frac{L_{p_{\mathrm{R}}}^{\lvert \ell_{\mathrm{R}} \rvert} \left( \rho^{2} \right)}{L_{p_{\mathrm{L}}}^{\lvert \ell_{\mathrm{L}} \rvert} \left( \rho^{2} \right)},
	\end{aligned}
\end{align}
the azimuthal index difference
\begin{align}
	m = \ell_{\mathrm{R}} - \ell_{\mathrm{L}},
\end{align}
and the offset phase
\begin{align}
	\Gamma \left( z \right) = -\gamma - \left( N_{\mathrm{R}} - N_{\mathrm{L}} \right) \varPhi \left( z \right) + \pi \left( p_{\mathrm{R}} - p_{\mathrm{L}} \right),
\end{align}
defined modulo $2 \pi$, with total mode numbers $N_{j} = N_{p_{j}, \ell_{j}}$ for polarization labels $j \in \{ \mathrm{L}, \mathrm{R} \}$.
The modal ratio has modulus $\lvert A(\rho) \rvert$ and an additional phase $\pi$ where the radial factor $A(\rho)$ is negative.
The signed radial factor includes the scaled radius $\rho$ raised to the difference in absolute azimuthal indices
\begin{align}
	\Delta \lvert \ell \rvert = \lvert \ell_{\mathrm{R}} \rvert - \lvert \ell_{\mathrm{L}} \rvert,
\end{align}
and the relative Gouy phase is proportional to the difference in total mode numbers
\begin{align}
	N_{\mathrm{R}} - N_{\mathrm{L}} = 2 \left( p_{\mathrm{R}} - p_{\mathrm{L}} \right) + \Delta \lvert \ell \rvert.
\end{align}

\section{Boundary behavior}
\label{sec:Sec3}

We parametrize the polarization texture,
\begin{align}
	\begin{aligned}
		\bm{s} \left( \bm{r}, z \right) =&~ \left( \sin \theta \left( r, z \right) \cos \chi \left( \bm{r}, z \right), \right. \\
		& \sin \theta \left( r, z \right) \sin \chi \left( \bm{r}, z \right), \left. \cos \theta \left( r, z \right) \right),
	\end{aligned}
\end{align}
with Stokes polar angle,
\begin{align}
	\cos \theta \left( r, z \right) = \frac{1 - \lvert \mu \left( r, z \right) \rvert^{2}}{1 + \lvert \mu \left( r, z \right) \rvert^{2}},
\end{align}
and Stokes azimuth angle,
\begin{align}
	\chi \left( \bm{r}, z \right) = m \phi + \Gamma \left( z \right) + \arg A \left( \rho \right),
\end{align}
to calculate the polarization texture charge as the signed spherical area covered, divided by the sphere's total area~\cite{McWilliam2023p23030155},
\begin{align}
	Q = \frac{1}{4 \pi} \int_{\mathbb{R}^{2}} \mathrm{d} x \, \mathrm{d} y \, \bm{s} \cdot \left( \partial_{x} \bm{s} \times \partial_{y} \bm{s} \right).
\end{align}
At dark rings we cancel the common radial factor in the modal ratio to extend the polarization texture.
At a singular core we take the inner integration radius to zero after integration.
At a fixed propagation plane $z$, the polar angle $\theta$ depends only on the transverse radius $r$, while the azimuth angle $\chi$ has winding number $m$.
On each transverse circle, the Stokes vector traces the corresponding latitude $\lvert m \rvert$ times as the azimuth $\phi$ goes around the beam axis, with orientation determined by the sign of $m$.
We integrate the resulting triple product $-m \, \partial_{r} \cos \theta(r, z) / r$ over the azimuth and radius to obtain the boundary charge~\cite{Gao2020p053513,Ye2024p240411530},
\begin{align}
	Q = \frac{m}{2} \left[ \cos \theta \left( 0, z \right) - \cos \theta \left( \infty, z \right) \right].
\end{align}
For polar limits, the boundary charge has the polarity--vorticity form $Q = m P$, with polarity $P = \left[ \cos \theta \left( 0, z \right) - \cos \theta \left( \infty, z \right) \right] / 2$~\cite{Shen2022p296}.

The generalized Laguerre polynomials have nonzero values at the origin,
\begin{align}
	L_{p}^{\lvert \ell \rvert} \left( 0 \right) = \frac{\left( p + \lvert \ell \rvert \right)!}{p! \, \lvert \ell \rvert!},
\end{align}
and the ratio of the two Laguerre factors approaches a finite nonzero constant at the propagation axis.
The modulus of the modal ratio then scales as $\rho^{\Delta \lvert \ell \rvert}$ near the origin~\cite{Ye2024p240411530}.
For a positive difference in absolute azimuthal indices, the modal ratio vanishes near the axis, the left-circular component dominates, and the polarization approaches the north pole, with third Stokes component $\cos \theta(0, z) = 1$.
For a negative difference, the modal ratio diverges near the propagation axis, the right-circular component dominates, and the polarization approaches the south pole, with third Stokes component $\cos \theta(0, z) = -1$.
For equal absolute azimuthal indices $\lvert \ell \rvert = \lvert \ell_{\mathrm{L}} \rvert = \lvert \ell_{\mathrm{R}} \rvert$, we obtain a finite nonzero modulus and an intermediate core latitude,
\begin{align}
	\begin{aligned}
		\lvert \mu \left( 0, z \right) \rvert =&~ \sqrt{\frac{p_{\mathrm{L}}! \, \left( p_{\mathrm{R}} + \lvert \ell \rvert \right)!}{p_{\mathrm{R}}! \, \left( p_{\mathrm{L}} + \lvert \ell \rvert \right)!}}, \\
		\cos \theta \left( 0, z \right) =&~ \frac{p_{\mathrm{R}}! \, \left( p_{\mathrm{L}} + \lvert \ell \rvert \right)! - p_{\mathrm{L}}! \, \left( p_{\mathrm{R}} + \lvert \ell \rvert \right)!}{p_{\mathrm{R}}! \, \left( p_{\mathrm{L}} + \lvert \ell \rvert \right)! + p_{\mathrm{L}}! \, \left( p_{\mathrm{R}} + \lvert \ell \rvert \right)!}.
	\end{aligned}
\end{align}
The intermediate core has a single limit at zero Stokes winding.
At nonzero winding, the Stokes vector traces the limiting latitude as the azimuth $\phi$ varies, leaving an angular singularity at the origin.

At large scaled radius, we use the leading term of each generalized Laguerre polynomial,
\begin{align}
	L_{p}^{\lvert \ell \rvert} \left( \rho^{2} \right) \sim \left( -1 \right)^{p} \frac{\rho^{2 p}}{p!}.
\end{align}
The ratio of these leading terms scales as $\rho^{2 \left( p_{\mathrm{R}} - p_{\mathrm{L}} \right)}$.
We combine this leading radial power with $\rho^{\Delta \lvert \ell \rvert}$ to obtain the exponent $N_{\mathrm{R}} - N_{\mathrm{L}}$ in the asymptotic modulus of the modal ratio,
\begin{align}
	\lvert \mu \left( r, z \right) \rvert \sim \sqrt{\frac{p_{\mathrm{L}}! \, \left( p_{\mathrm{L}} + \lvert \ell_{\mathrm{L}} \rvert \right)!}{p_{\mathrm{R}}! \, \left( p_{\mathrm{R}} + \lvert \ell_{\mathrm{R}} \rvert \right)!}} \rho^{N_{\mathrm{R}} - N_{\mathrm{L}}}.
\end{align}
The sign of the radial exponent $N_{\mathrm{R}} - N_{\mathrm{L}}$ determines the asymptotic class.
For a positive difference, the modal ratio diverges and the polarization approaches the south pole, with third Stokes component $\cos \theta(\infty, z) = -1$.
For a negative difference, the modal ratio vanishes and the polarization approaches the north pole, with third Stokes component $\cos \theta(\infty, z) = 1$.
For equal total mode numbers $N_{\mathrm{R}} = N_{\mathrm{L}}$, we obtain a finite nonzero modulus and an intermediate asymptotic latitude,
\begin{align}
	\begin{aligned}
		\lvert \mu \left( \infty, z \right) \rvert =&~ \sqrt{\frac{p_{\mathrm{L}}! \, \left( p_{\mathrm{L}} + \lvert \ell_{\mathrm{L}} \rvert \right)!}{p_{\mathrm{R}}! \, \left( p_{\mathrm{R}} + \lvert \ell_{\mathrm{R}} \rvert \right)!}}, \\
		\cos \theta \left( \infty, z \right) =&~ \frac{p_{\mathrm{R}}! \, \left( p_{\mathrm{R}} + \lvert \ell_{\mathrm{R}} \rvert \right)! - p_{\mathrm{L}}! \, \left( p_{\mathrm{L}} + \lvert \ell_{\mathrm{L}} \rvert \right)!}{p_{\mathrm{R}}! \, \left( p_{\mathrm{R}} + \lvert \ell_{\mathrm{R}} \rvert \right)! + p_{\mathrm{L}}! \, \left( p_{\mathrm{L}} + \lvert \ell_{\mathrm{L}} \rvert \right)!}.
	\end{aligned}
\end{align}
At nonzero Stokes winding, the asymptotic polarization approaches different points on this latitude as the transverse azimuth $\phi$ varies.
With zero winding and equal total mode numbers, the two scalar modes coincide and the polarization is uniform.

\begin{figure}[t]
    \centering
    \includegraphics[width=0.75 \textwidth]{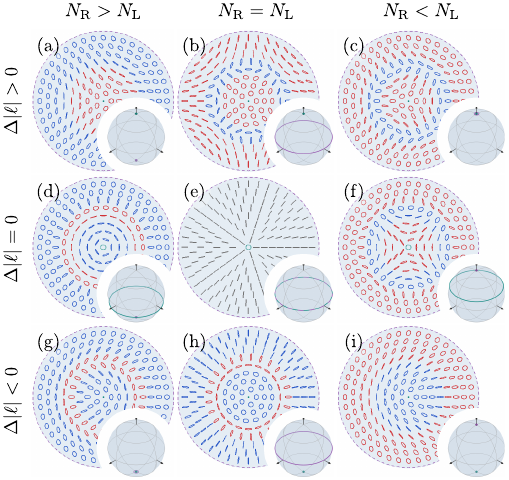}
    \caption{Polarization textures for the nine core and asymptotic classes of Table~\ref{tab:Table1}, with their boundary states on the inset Poincar\'e spheres.
    Teal marks the core limit as $r\to0$, and violet marks the asymptotic limit as $r\to\infty$.}
    \label{fig:Fig2}
\end{figure}

A polarization texture is open at a boundary when its limiting state depends on the direction of approach.
A continuous map from the compactified transverse plane to the Poincar\'e sphere requires a single polarization limit at each boundary.
With zero Stokes winding or two polar limits, the charge $Q$ is the degree of the extended map.
At a direction-dependent boundary, the winding and limiting latitudes still determine the integrated charge, although the texture has no degree on the compactified plane.
We combine the core and asymptotic classes in Table~\ref{tab:Table1} and show their nine combinations in Fig.~\ref{fig:Fig2}.

\begin{table*}[t]
\caption{
	Core and asymptotic latitudes of the two-mode Laguerre--Gaussian polarization family.
	At nonzero Stokes winding number $m$, an intermediate latitude has an azimuth-dependent limiting direction.
	At zero winding both limits are single valued and the charge vanishes.
}
\label{tab:Table1}
\begin{ruledtabular}
\begin{tabular}{cccc}
\shortstack{Core \\ class}
& \shortstack{$N_{\mathrm{R}} > N_{\mathrm{L}}$ \\ $\theta(\infty) = \pi$}
& \shortstack{$N_{\mathrm{R}} = N_{\mathrm{L}}$ \\ $0 < \theta(\infty) < \pi$}
& \shortstack{$N_{\mathrm{R}} < N_{\mathrm{L}}$ \\ $\theta(\infty) = 0$} \\[2pt]
\hline
\noalign{\vskip 2pt}
\shortstack{$\Delta\lvert\ell\rvert > 0$ \\ $\theta(0) = 0$}
& \shortstack{skyrmion \\ $Q=m$ \\ Fig.~\ref{fig:Fig5}}
& \shortstack{open \\ $Q=(m/2)[1-\cos\theta(\infty)]$ \\ Fig.~\ref{fig:Fig6}}
& \shortstack{charge-neutral \\ $Q=0$ \\ Fig.~\ref{fig:Fig7}} \\[6pt]
\shortstack{$\Delta\lvert\ell\rvert = 0$ \\ $0 < \theta(0) < \pi$}
& \shortstack{intermediate-core \\ $Q=(m/2)[1+\cos\theta(0)]$ \\ Fig.~\ref{fig:Fig8}}
& \shortstack{equatorial \\ $Q=0$ \\ Fig.~\ref{fig:Fig9}}
& \shortstack{intermediate-core \\ $Q=(m/2)[\cos\theta(0)-1]$ \\ \phantom{Fig.~9}} \\[6pt]
\shortstack{$\Delta\lvert\ell\rvert < 0$ \\ $\theta(0) = \pi$}
& \shortstack{charge-neutral \\ $Q=0$ \\ \phantom{Fig.~9}}
& \shortstack{open \\ $Q=-(m/2)[1+\cos\theta(\infty)]$ \\ \phantom{Fig.~9}}
& \shortstack{south-core skyrmion \\ $Q=-m$ \\ \phantom{Fig.~9}} \\[2pt]
\end{tabular}
\end{ruledtabular}
\end{table*}

The opposite-pole sectors have skyrmion charge magnitude $\lvert Q \rvert = \lvert m \rvert$, while the equal-pole sectors have zero charge.
A polar core requires unequal absolute azimuthal indices and nonzero Stokes winding.
A polar core and intermediate asymptotic latitude require an even, nonzero difference in absolute azimuthal indices $\Delta \lvert \ell \rvert = 2 (p_{\mathrm{L}} - p_{\mathrm{R}})$.

For an intermediate core latitude, equal azimuthal indices imply zero Stokes winding.
Opposite nonzero azimuthal indices imply the index difference $m = 2 \ell_{\mathrm{R}}$ and a singular core, including textures with a polar asymptote~\cite{Ye2024p240411530}.
In the equatorial sector the radial factors coincide, and the polarization texture lies on the equator of linear states, with zero charge.

\begin{figure}[t]
    \centering
    \includegraphics[width=0.75 \textwidth]{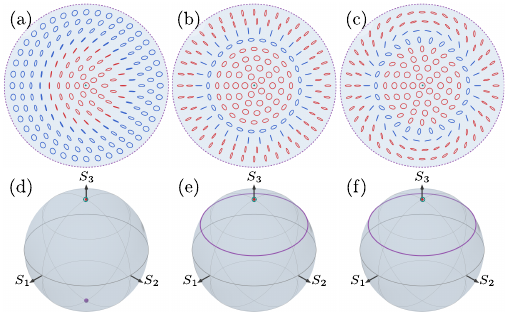}
    \caption{Polarization ellipses (a)--(c) and the corresponding core and asymptotic states on Poincar\'e spheres (d)--(f) for three pairs of left- and right-circular modes at $z=0$ and $\gamma=0$.
    The pairs are (a,d) $\mathrm{LG}_{0,0}$ and $\mathrm{LG}_{0,1}$ with $Q=1$, (b,e) $\mathrm{LG}_{1,1}$ and $\mathrm{LG}_{0,3}$ with $Q=1/2$, and (c,f) $\mathrm{LG}_{1,-1}$ and $\mathrm{LG}_{0,3}$ with $Q=1$.
    Teal marks the core limit as $r\to0$, and violet marks the asymptotic limit as $r\to\infty$.
    The first pair has a single asymptotic polarization state; the violet sphere circles trace the direction-dependent asymptotic states of the other two pairs.}
    \label{fig:Fig3}
\end{figure}

We show the polarization ellipses of left-circular $\mathrm{LG}_{0,0}$ and right-circular $\mathrm{LG}_{0,1}$ in Fig.~\ref{fig:Fig3}(a), those of left-circular $\mathrm{LG}_{1,1}$ and right-circular $\mathrm{LG}_{0,3}$ in Fig.~\ref{fig:Fig3}(b), those of left-circular $\mathrm{LG}_{1,-1}$ and right-circular $\mathrm{LG}_{0,3}$ in Fig.~\ref{fig:Fig3}(c), the corresponding teal core and violet asymptotic points on the Poincar\'e sphere in Fig.~\ref{fig:Fig3}(d), and the teal core points and violet asymptotic circles in Fig.~\ref{fig:Fig3}(e) and Fig.~\ref{fig:Fig3}(f).
The open pairs share the asymptotic latitude $\cos\theta(\infty,z)=1/2$, but their windings $m=2$ and $m=4$ correspond to charges $Q=1/2$ and $Q=1$.

\section{Interior morphology}
\label{sec:Sec4}

We locate the pole and dark rings at the positive zeros of the right and left Laguerre factors to calculate the charge of each concentric region,
\begin{align}
	\begin{aligned}
		L_{p_{\mathrm{R}}}^{\lvert \ell_{\mathrm{R}} \rvert} \left( \rho^{2} \right) =&~ 0, \\
		L_{p_{\mathrm{L}}}^{\lvert \ell_{\mathrm{L}} \rvert} \left( \rho^{2} \right) =&~ 0,
	\end{aligned}
\end{align}
with scaled ring radii
\begin{align}
	\rho_{j, 1} < \rho_{j, 2} < \cdots < \rho_{j, p_{j}},
\end{align}
for polarization labels $j \in \{\mathrm{L}, \mathrm{R}\}$ with positive radial index $p_{j} > 0$, since each degree-$p_{j}$ polynomial has $p_{j}$ positive simple zeros~\cite{NISTDLMFLaguerre}.
At a root of only the right component, the field is left-circular and the ring maps to the north pole.
At a root of only the left component, the field is right-circular and the ring maps to the south pole.
At a shared root both components vanish and the ring is dark.
We cancel the common simple factor to obtain a finite, nonzero modal ratio and extend the polarization texture smoothly across the ring.

With $n_{\mathrm{c}}$ shared roots, the texture has $p_{\mathrm{R}} - n_{\mathrm{c}}$ north-pole rings, $p_{\mathrm{L}} - n_{\mathrm{c}}$ south-pole rings, and $n_{\mathrm{c}}$ dark rings.
We order the $K = p_{\mathrm{L}} + p_{\mathrm{R}} - n_{\mathrm{c}}$ distinct scaled ring radii,
\begin{align}
	\rho_{0} < \rho_{1} < \cdots < \rho_{K} < \rho_{K+1},
\end{align}
with inner boundary $\rho_{0} = 0$ and outer limit $\rho_{K+1} = \infty$.
At a propagation plane $z$, we integrate between successive transverse radii $r_{n}$ and $r_{n+1}$ to calculate the annular charge,
\begin{align}
	Q_{n} = \frac{m}{2} \left[ \cos \theta \left( r_{n}, z \right) - \cos \theta \left( r_{n+1}, z \right) \right],
\end{align}
for annulus indices $n = 0, 1, \ldots, K$.
At a dark ring we use the limiting latitude of the extended polarization texture.
An annulus bounded by two rings at the same pole has zero net charge.
The annulus between a north-pole inner ring and a south-pole outer ring has charge $Q_{n} = m$; reversing the pole order changes its charge to $Q_{n} = -m$.
Summing the annular charges cancels the internal boundary terms pairwise and recovers the full-plane charge.

The distinction between the full-plane charge and its annular contributions is clear when we vary the shared radial index $p = 0, 1, \ldots$ of left-circular $\mathrm{LG}_{p,0}$ and right-circular $\mathrm{LG}_{p,-1}$ modes.
The pair at $p = 0$ forms a simple paraxial skyrmion without finite-radius pole rings~\cite{Shen2022p296,Cisowski2023p2200350}.
For every radial index $p$, the polarization approaches the north pole at the core and the south pole at infinity, with Stokes winding $m = -1$ and total charge $Q = -1$.
We establish the ring order from the Laguerre identity~\cite{NISTDLMFLaguerre},
\begin{align}
	L_{p}^{1} \left( u \right) = L_{p}^{0} \left( u \right) - \frac{\mathrm{d} L_{p}^{0} \left( u \right)}{\mathrm{d} u},
\end{align}
with squared scaled radius $u = \rho^{2}$.
At each simple zero of the left Laguerre factor $L_{p}^{0}$, the right factor equals the negative of the left factor's nonzero slope, and the ring maps to the south pole.
The slope alternates in sign at consecutive zeros of the left factor, placing at least one zero of the right factor between each pair.
At the last zero of the left factor, the slope has the sign of the common leading coefficient.
At that zero, the right factor has the opposite sign from its sign at large radius and must cross zero farther out.
The $p - 1$ intervening zeros and the outer zero exhaust the $p$ zeros of the right factor.
The scaled ring radii interlace,
\begin{align}
	\rho_{\mathrm{L}, 1} < \rho_{\mathrm{R}, 1} < \cdots < \rho_{\mathrm{L}, p} < \rho_{\mathrm{R}, p},
\end{align}
for positive radial index $p > 0$.
These $2 p$ alternating pole rings begin with a south-pole ring and end with a north-pole ring.
The $2p$ rings, north-pole core, and south-pole asymptote bound $2p+1$ regions whose polarization runs alternately north to south and south to north.
We calculate the annular charge from the alternating latitudes and azimuthal index difference $m = -1$,
\begin{align}
	Q_{n} = \left( -1 \right)^{n+1},
\end{align}
for annulus indices $n = 0, 1, \ldots, 2p$.
The $p$ positive annular charges cancel $p$ of the $p + 1$ negative charges, leaving total charge $Q = -1$; raising the shared radial index by one adds a region of each sign.

\section{Propagation behavior}
\label{sec:Sec5}

At fixed scaled radius $\rho$ and azimuth $\phi$, the modal ratio retains its radial factor and acquires the relative Gouy phase $-(N_{\mathrm{R}} - N_{\mathrm{L}}) \varPhi(z)$.
We relate the normalized Stokes components at plane $z$ to their values at the waist,
\begin{align}
	\begin{aligned}
		s_{1} \left( \bm{r}, z \right) + i s_{2} \left( \bm{r}, z \right) =&~ e^{-i \left( N_{\mathrm{R}} - N_{\mathrm{L}} \right) \varPhi \left( z \right)} \\
		& \times \left[ s_{1} \left( \bm{r}^{\prime}, 0 \right) + i s_{2} \left( \bm{r}^{\prime}, 0 \right) \right], \\
		s_{3} \left( \bm{r}, z \right) =&~ s_{3} \left( \bm{r}^{\prime}, 0 \right),
	\end{aligned}
\end{align}
with corresponding waist coordinate $\bm{r}^{\prime} = w_{0} \bm{r} / w(z)$.
The ring radii scale with the beam width $w(z)$ and keep their order.
We find that free propagation preserves each annular charge $Q_{n}$ and the full-plane charge $Q$ of both closed and open textures in this common-waist family because the edge latitudes and Stokes winding number $m$ stay fixed.
When the modes differ in waist size or waist plane, their beam widths can cross during propagation, changing the asymptotic polarization and, at nonzero Stokes winding, the charge~\cite{Gao2020p053513,Ye2024p240411530}.

\begin{figure}[t]
    \centering
    \includegraphics[width=0.75 \textwidth]{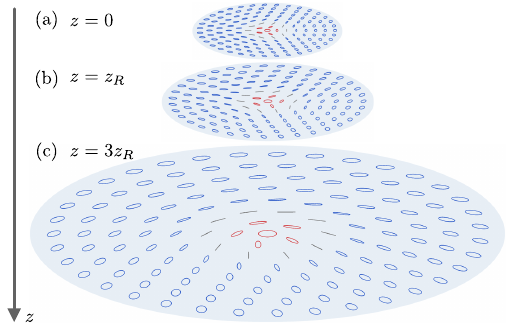}
    \caption{Free propagation of left-circular $\mathrm{LG}_{0,0}$ and right-circular $\mathrm{LG}_{0,-1}$ modes, showing expansion of the Gaussian beam width $w(z)$ and Gouy-phase rotation of the polarization ellipses.
    (a) $\varPhi(0)=0$, (b) $\varPhi(z_R)=\pi/4$, and (c) $\varPhi(3z_R)=\arctan 3$.
    Where defined, the ellipse axes turn by $-\varPhi(z)/2$ at corresponding scaled positions for phase parameter $\gamma=0$.
    The arrow marks increasing $z$ with schematic plane spacing.}
    \label{fig:Fig4}
\end{figure}

For equal total mode numbers, the relative Gouy phase vanishes and the polarization texture repeats under transverse dilation, as in fixed-order total angular momentum coherent state fields~\cite{AguirreOlivas2026p3221}.
For unequal total mode numbers, the Stokes azimuth angle $\chi$ changes by the relative Gouy phase while the polar angle $\theta$ stays fixed.
The local ellipse axis rotates by half that angle wherever it has a well-defined direction.
At fixed scaled radius $\rho$ and nonzero Stokes winding number $m$, the polarization state at azimuth $\phi$ matches the waist-plane state at azimuth $\phi-(N_{\mathrm{R}}-N_{\mathrm{L}})\varPhi(z)/m$.
The texture turns spatially by $(N_{\mathrm{R}}-N_{\mathrm{L}})\varPhi(z)/m$, while the local ellipse axis turns by $-(N_{\mathrm{R}}-N_{\mathrm{L}})\varPhi(z)/2$ at a fixed scaled position.
The radially nodeless full Poincar\'e beam is an established example of this rotation~\cite{Beckley2010p10777}.
For left-circular $\mathrm{LG}_{p,0}$ and right-circular $\mathrm{LG}_{p,-1}$, the total mode numbers are $N_{\mathrm{L}} = 2 p + 1$ and $N_{\mathrm{R}} = 2 p + 2$, and the Stokes azimuth rotates by $-\varPhi(z)$ for every radial index $p$.
We show the polarization ellipses at $z=0$ in Fig.~\ref{fig:Fig4}(a), at $z=z_{R}$ in Fig.~\ref{fig:Fig4}(b), and at $z=3z_{R}$ in Fig.~\ref{fig:Fig4}(c) for the radially nodeless pair, illustrating the transverse expansion and rotation.

\section{Polarization textures by boundary class}
\label{sec:Sec6}

\begin{figure}[t]
    \centering
    \includegraphics[width=0.75 \textwidth]{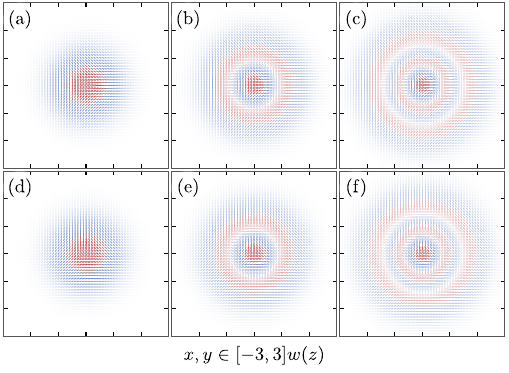}
    \caption{
        Polarization ellipses for left-circular $\mathrm{LG}_{p,0}$ and right-circular $\mathrm{LG}_{p,-1}$ modes, with (a,d) $p = 0$, (b,e) $p = 1$, and (c,f) $p = 2$; all three pairs have charge $Q = -1$.
        The upper row (a)--(c) shows the waist plane $z = 0$ and the lower row (d)--(f) shows the propagated plane $z = 3 z_{R}$, both at phase parameter $\gamma = 0$, with transverse coordinates $x / w(z)$ and $y / w(z)$.
        Red marks left-handed polarization, blue marks right-handed polarization, and black marks linearly polarized states; ellipse opacity is the square root of the ratio of local intensity $S_{0}$ to its sampled peak in each map.
    }
    \label{fig:Fig5}
\end{figure}

\begin{figure}[t]
    \centering
    \includegraphics[width=0.75 \textwidth]{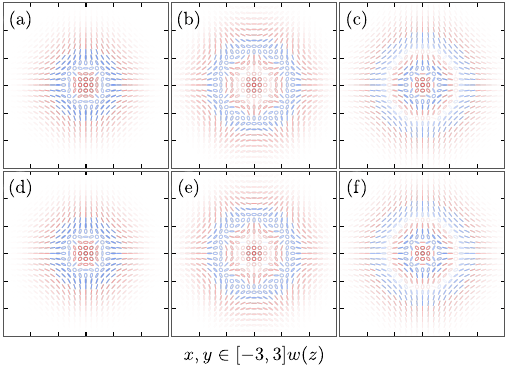}
    \caption{
        Polarization maps with a north-pole core and an intermediate asymptotic latitude, using the plotting conventions of Fig.~\ref{fig:Fig5}.
        The LG mode pairs, listed left-circular first and right-circular second, are (a,d) $\mathrm{LG}_{1,0}$ and $\mathrm{LG}_{0,-2}$ with one south-pole ring and charge $Q = -2/3$, (b,e) $\mathrm{LG}_{2,0}$ and $\mathrm{LG}_{0,-4}$ with two south-pole rings and charge $Q = -4/7$, and (c,f) $\mathrm{LG}_{2,0}$ and $\mathrm{LG}_{1,-2}$ with mixed pole rings and charge $Q = -4/5$.
    }
    \label{fig:Fig6}
\end{figure}

\begin{figure}[t]
    \centering
    \includegraphics[width=0.75 \textwidth]{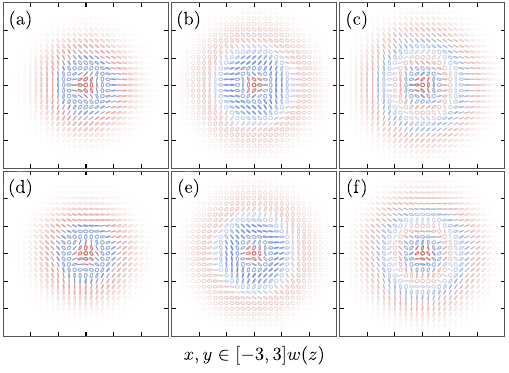}
    \caption{
        Polarization maps with a north-pole core and a north-pole asymptote, using the plotting conventions of Fig.~\ref{fig:Fig5}.
        The LG mode pairs, listed left-circular first and right-circular second, are (a,d) $\mathrm{LG}_{1,0}$ and $\mathrm{LG}_{0,-1}$ with one south-pole ring, (b,e) $\mathrm{LG}_{2,0}$ and $\mathrm{LG}_{0,-1}$ with two south-pole rings, and (c,f) $\mathrm{LG}_{2,0}$ and $\mathrm{LG}_{1,-1}$ with mixed pole rings; all three pairs have charge $Q = 0$.
    }
    \label{fig:Fig7}
\end{figure}

\begin{figure}[t]
    \centering
    \includegraphics[width=0.75 \textwidth]{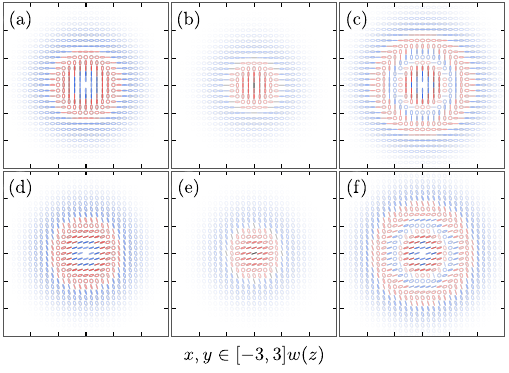}
    \caption{
        Polarization maps with an intermediate core latitude and a south-pole asymptote, using the plotting conventions of Fig.~\ref{fig:Fig5}.
        The LG mode pairs, listed left-circular first and right-circular second, are (a,d) $\mathrm{LG}_{0,-1}$ and $\mathrm{LG}_{1,-1}$ with one north-pole ring, (b,e) $\mathrm{LG}_{0,0}$ and $\mathrm{LG}_{1,0}$ with one north-pole ring, and (c,f) $\mathrm{LG}_{1,-1}$ and $\mathrm{LG}_{2,-1}$ with mixed pole rings; all three pairs have charge $Q = 0$.
    }
    \label{fig:Fig8}
\end{figure}

\begin{figure}[t]
    \centering
    \includegraphics[width=0.75 \textwidth]{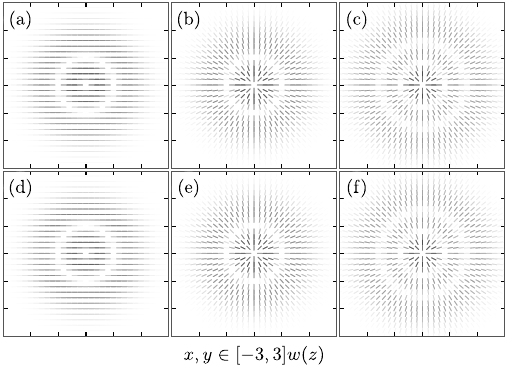}
    \caption{
        Polarization maps for the equatorial sector, using the plotting conventions of Fig.~\ref{fig:Fig5}.
        The LG mode pairs, listed left-circular first and right-circular second, are (a,d) $\mathrm{LG}_{1,-1}$ and $\mathrm{LG}_{1,-1}$ with uniform polarization, (b,e) $\mathrm{LG}_{1,-1}$ and $\mathrm{LG}_{1,1}$ with one dark ring in a vector vortex texture, and (c,f) $\mathrm{LG}_{2,-1}$ and $\mathrm{LG}_{2,1}$ with two dark rings in a vector vortex texture; all three pairs have charge $Q = 0$.
    }
    \label{fig:Fig9}
\end{figure}

We illustrate five of our nine boundary classes in Figs.~\ref{fig:Fig5}--\ref{fig:Fig9} and obtain the other four by exchanging the left- and right-circular modes, which reverses the boundary latitudes, Stokes winding, and Gouy rotation while preserving the charge.

\textbf{North core, south asymptote.}
Left-circular $\mathrm{LG}_{0,0}$ and right-circular $\mathrm{LG}_{0,m}$ modes with nonzero azimuthal index difference $m$ form the standard skyrmions with a north-pole core and south-pole asymptote~\cite{Gao2020p053513,Gao2021p049901,Shen2022p296,Cisowski2023p2200350}.
For this boundary class, the charge is $Q = m = \ell_{\mathrm{R}} - \ell_{\mathrm{L}}$ even when the modes have radial nodes.
We show left-circular $\mathrm{LG}_{p,0}$ and right-circular $\mathrm{LG}_{p,-1}$ at $p = 0$ in Fig.~\ref{fig:Fig5}(a) and Fig.~\ref{fig:Fig5}(d), at $p = 1$ in Fig.~\ref{fig:Fig5}(b) and Fig.~\ref{fig:Fig5}(e), and at $p = 2$ in Fig.~\ref{fig:Fig5}(c) and Fig.~\ref{fig:Fig5}(f).
Each increase in $p$ adds a south-pole ring and a north-pole ring whose annular charges cancel; all three textures have charge $Q = -1$ and Stokes rotation angle $-\varPhi(z)$.

\textbf{North core, intermediate asymptote.}
These open textures have a north-pole core and an azimuth-dependent polarization at infinity.
We show left-circular $\mathrm{LG}_{1,0}$ and right-circular $\mathrm{LG}_{0,-2}$ with one south-pole ring and charge $Q = -2/3$ in Fig.~\ref{fig:Fig6}(a) and Fig.~\ref{fig:Fig6}(d), left-circular $\mathrm{LG}_{2,0}$ and right-circular $\mathrm{LG}_{0,-4}$ with two south-pole rings and charge $Q = -4/7$ in Fig.~\ref{fig:Fig6}(b) and Fig.~\ref{fig:Fig6}(e), and left-circular $\mathrm{LG}_{2,0}$ and right-circular $\mathrm{LG}_{1,-2}$ with both types of pole ring and charge $Q = -4/5$ in Fig.~\ref{fig:Fig6}(c) and Fig.~\ref{fig:Fig6}(f).
Their different Stokes windings and finite modal ratios at infinity account for the different charges; because the total mode numbers are equal, the polarization texture repeats in scaled coordinates during propagation.

\textbf{North core, north asymptote.}
The polarization reaches the south pole at finite radii but approaches the north pole at both boundaries, so its annular charges cancel and the total charge vanishes.
We show a one-ring skyrmionium from left-circular $\mathrm{LG}_{1,0}$ and right-circular $\mathrm{LG}_{0,-1}$ in Fig.~\ref{fig:Fig7}(a) and Fig.~\ref{fig:Fig7}(d); its radial partition also appears in earlier Poincar\'e beams~\cite{Galvez2012p2925}.
Left-circular $\mathrm{LG}_{2,0}$ and right-circular $\mathrm{LG}_{0,-1}$ have two south-pole rings in Fig.~\ref{fig:Fig7}(b) and Fig.~\ref{fig:Fig7}(e), while left-circular $\mathrm{LG}_{2,0}$ and right-circular $\mathrm{LG}_{1,-1}$ have a south-, north-, south-pole ring sequence in Fig.~\ref{fig:Fig7}(c) and Fig.~\ref{fig:Fig7}(f).
The Stokes vectors rotate by $+\varPhi(z)$, $+3\varPhi(z)$, and $+\varPhi(z)$ at corresponding scaled positions in the three columns.

\textbf{Intermediate core, south asymptote.}
For equal signed azimuthal indices, the Stokes winding and charge both vanish despite the different ring patterns.
We show left-circular $\mathrm{LG}_{0,-1}$ and right-circular $\mathrm{LG}_{1,-1}$ with one north-pole ring in Fig.~\ref{fig:Fig8}(a) and Fig.~\ref{fig:Fig8}(d), left-circular $\mathrm{LG}_{0,0}$ and right-circular $\mathrm{LG}_{1,0}$ with one north-pole ring in Fig.~\ref{fig:Fig8}(b) and Fig.~\ref{fig:Fig8}(e), and left-circular $\mathrm{LG}_{1,-1}$ and right-circular $\mathrm{LG}_{2,-1}$ with a north-, south-, north-pole ring sequence in Fig.~\ref{fig:Fig8}(c) and Fig.~\ref{fig:Fig8}(f).
The first and third pairs have zero intensity at the core but a direction-independent polarization limit; the second has nonzero core intensity.
All three pairs have a total mode number difference of two, so their Stokes vectors rotate by $-2\varPhi(z)$ at corresponding scaled positions.

\textbf{Equatorial sector.}
Equal total mode numbers and equal absolute azimuthal indices imply equal radial indices $p_{\mathrm{R}} = p_{\mathrm{L}}$.
The modal ratio has unit modulus because the radial factors coincide, and the polarization is linear wherever the intensity is nonzero.
The polarization texture extends across the shared dark rings on the equator, and its charge vanishes.
For equal azimuthal indices, the modal ratio is constant, $\mu = e^{-i \gamma}$, and the polarization is uniform, horizontal at phase parameter $\gamma = 0$.
For opposite nonzero azimuthal indices, the modal ratio is $\mu = e^{i (2 \ell_{\mathrm{R}} \phi - \gamma)}$, and the vector vortex texture has ellipse-axis angle $\ell_{\mathrm{R}} \phi - \gamma / 2$ modulo $\pi$~\cite{Galvez2012p2925}.
Its core is singular and its asymptotic polarization depends on the azimuth.
For azimuthal indices $\ell_{\mathrm{L}} = -1$ and $\ell_{\mathrm{R}} = 1$, the phase parameter $\gamma = 0$ corresponds to a radial vector beam and $\gamma = \pi$ to an azimuthal vector beam~\cite{Zhan2009p1}.
We show uniform polarization from left-circular and right-circular $\mathrm{LG}_{1,-1}$ in Fig.~\ref{fig:Fig9}(a) and Fig.~\ref{fig:Fig9}(d), a vector vortex texture from left-circular $\mathrm{LG}_{1,-1}$ and right-circular $\mathrm{LG}_{1,1}$ with one dark ring in Fig.~\ref{fig:Fig9}(b) and Fig.~\ref{fig:Fig9}(e), and one from left-circular $\mathrm{LG}_{2,-1}$ and right-circular $\mathrm{LG}_{2,1}$ with two dark rings in Fig.~\ref{fig:Fig9}(c) and Fig.~\ref{fig:Fig9}(f).
The opposite-index pairs have the same Stokes winding and no Gouy rotation despite the additional dark ring.
Across our boundary classes, equal boundary states and total charge can accompany different pole-ring sequences and Gouy rotations.

\section{Conclusion}
\label{sec:Sec7}

We classified nine boundary classes for two equally weighted Laguerre--Gaussian modes with opposite circular polarizations and a common waist from the difference in absolute azimuthal indices at the core and the difference in total mode numbers at infinity.
At nonzero Stokes winding, an intermediate latitude at either boundary makes the polarization texture open and can accompany noninteger integrated charge.
With single-valued limits at both boundaries, the charge is the integer degree of the extended polarization texture.

We located pole and dark rings from the Laguerre roots and calculated each annular charge from its edge latitudes and the Stokes winding.
Free propagation scales the rings and preserves both the annular and full-plane charges; unequal total mode numbers rotate the Stokes vectors through the relative Gouy phase.
For left-circular modes of azimuthal index zero paired with right-circular modes of index minus one, each increment in their shared radial index adds a north-pole and a south-pole ring with canceling annular charges.
The total charge stays at minus one, and the Stokes rotation angle equals the negative of the Gouy phase for every radial index.

Equal boundary states and total charge can accompany different pole-ring sequences and Gouy rotations.
We can use the mode indices to choose beams with a desired polarization pattern and charge at a chosen plane.
Our~approach may guide the design of light fields structured in intensity, phase, and polarization for optical communication, imaging, and the manipulation of matter.


\section*{Acknowledgements}
L.~E.~L.~R. thanks Carmelo Rosales-Guzm\'an for his guidance as her supervisor.
B.~M.~R.~L. acknowledges the support and hospitality he receives as an affiliate visiting colleague in the Department of Physics and Astronomy at the University of New Mexico and thanks Benjam\'in de Jes\'us P\'erez Garc\'ia for fruitful discussions.

\section*{Funding}
This work received no funding.

\section*{Data Availability Statement}
The data that support the findings of this study are available from the corresponding author upon reasonable request.
The corresponding author can also provide the figure-generation and verification code.

\section*{Disclosures}
The authors declare no conflicts of interest.

\section*{Tool and Computational Resource}
Claude Science with the OpenAlex API was used for literature searches.
OpenAI GPT-6 Sol and Claude Opus 5.5 were used as adversarial reviewers to improve the clarity of our exposition and our use of language.
We take full responsibility for the accuracy, integrity, and attribution of the final work.



%

\end{document}